\documentclass{article}
\usepackage{frascatiphys}
\usepackage{graphicx}
\usepackage{lineno}
\begin{document}
\title{ ARE COSMIC RAYS STILL A VALUABLE PROBE OF
LORENTZ INVARIANCE VIOLATIONS IN THE AUGER ERA?
}
\pagenumbering{gobble}
%\linenumbers\relax
\author{
Roberto Aloisio       \\
{\em INAF Arcetri,Firenze, Italy and Gran Sasso Science Institute (INFN), L'Aquila, Italy} \\
Denise Boncioli, Aurelio F. Grillo        \\
{\em INFN Laboratori Nazionali del Gran Sasso, Assergi , Italy}
\\
Piera L. Ghia        \\
{\em CNRS/IN2P3 - LPNHE Paris} \\
Armando di Matteo,        \\
{\em INFN and Physics Department, Universit\`a di L'Aquila, L'Aquila, Italy} \\
 Sergio Petrera        \\
{\em Physics Department, Universit\`a di L'Aquila, INFN and Gran Sasso Science Institute, L'Aquila, Italy} \\
Francesco Salamida        \\
{\em CNRS/IN2P3 - IPN Orsay}
}
\maketitle
\baselineskip=11.6pt
\begin{abstract}
Relativistic Invariance might be modified by Quantum Gravity effects. The interesting point which emerged in the last fifteen years is that remnants of possible Lorentz Invariance Violations could  be present at energies much lower than their natural scale, and possibly affect  Ultra High Energy Cosmic Rays phenomena. We discuss their status in  the view of recent data from the Pierre Auger Observatory. 
\end{abstract}
\baselineskip=14pt

\section{Introduction \label{sec.intro}}
Relativistic Invariance is the fundamental space-time symmetry.
If General Relativity and Quantum Mechanics can be reconciled, space-time could be subject to quantum fluctuations and the Lorentz Invariant space-time could emerge as a semiclassical limit of Quantum Gravity (QG). Lorentz Invariance Violations (LIV) can therefore be possible. Although these effects may only be very small, it has been shown in the last two decades that measurable effects can  be present even at energies much lower than the Quantum Gravity scale. In particular possible LIV effects could show themselves in Ultra High Energy Cosmic Rays (UHECR) phenomena. \\
The possibility of putting extremely strong limits on, at least some, LIV parameters from UHECRs detection was firstly quantitatively discussed in\cite{aletal} and later on refined in many ways. Consequently, as soon as the evidence of the suppression in the spectrum of UHECRs around $5 \cdot 10^{19}~ eV$ became undisputable, based on results from HiRes\cite{Hires} and Auger\cite{fluxpao}, limits on those violating parameter were derived. A discussion and references can be found in\cite{lib013}.\\
Here we discuss the status of these bounds in the light of recent interpretation of measurements by the Pierre Auger Observatory\cite{ICRC2013} (see e.g. Aloisio 2013\cite{abb}) for which the observed suppression in the spectrum might be due to the maximum cosmic ray acceleration energy at the sources rather than to an effect of their propagation in extra-galactic space.
%\input{intro.tex}
%\section{Lorentz Invariance Violations: generalities\label{sec.livg}}
\section{Lorentz Invariance Violations: effects on UHECR propagation \label{sec.livprop}}
The aim of this paper is purely phenomenological and a general discussion of LI violating terms that can affect UHECR physics is out of its scope\cite{lib013}\cite{Kost14}. To parametrize departures from relativistic invariance we follow here the  approach of\cite{aletal},
which amounts to assuming that the relation, connecting the energy and momentum of a particle (dispersion relation), is modified as:
\begin{equation}
E_i^2-p_i^2=m_i^2 \Rightarrow \mu_i(E,p,m_P)  \approx m_i^2+{\frac{f_i}{ m_P^n}} E_i^{2+n}
\label{eq3}
\end{equation}
where $p=\vert \overrightarrow{p} \vert $, $\mu$ is an arbitrary function of momenta and energy, $ m_P \approx 2 \cdot 10^{28}~ eV$ is the possible scale where QG effects become important and $f_i$, which can have both signs,  parametrizes the strength of LIV for particle $i$. The last equality reflects the fact that LI is an exceedingly good approximation of the physics we know, so that modifications are expected to be quite small, making an expansion of the LIV dispersion relation in terms of $1/m_P$ appropriate. In practical terms, only $n=1,2$ will be relevant\cite{aletal}.\\ 
The right hand side of eq.\ref{eq3} is invariant when $f=0$. 
We will assume normal conservation of energy and momentum. 
Finally we assume that, in nuclei, LIV only affects nucleons: this implies that, for a nucleus of atomic number $A$, effectively  $m_p \rightarrow A m_P$.
From  eq.\ref{eq3} it is clear that the correction term is always much smaller than both ($E^2,p^2$) even for  $E \approx 10^{20}~eV$. However, as soon as\footnote{Since at the leading LIV order $E \approx p$ we will use them without distinction.} $p \geq (m_i^2 m_P^n /|f_i|)^{1/(2+n)}$ the correction becomes larger than the mass of the particle, and this can lead to very important effects\cite{aletal}.
We consider here how LIV affects the threshold energy for the Greisen\cite{G}, Zatsepin, Kuzmin\cite{ZK}  process $p \gamma_{bkg} \rightarrow (p,n) \pi$, where $\gamma_{bkg}$ is a photon of Cosmic Microwave Background  or Infrared  radiation. The threshold for this process, in a LIV world, is modified:
\begin{equation}
E_{GZK}\approx \frac{m_p m_{\pi}}{2 \omega_{\gamma}} \Rightarrow
E_{GZK}\approx \frac{\mu(E_p,p_p,m_p,m_P) \mu(E_{\pi},p_{\pi},m_{\pi},m_P)}{2 \omega_{\gamma}}
\label{eq6}
\end{equation} 
($\omega_{\gamma}$ being the energy of the background photon). The last equation has to be solved for $E_p=E_{GZK}$.
For our simplified treatement, we will assume that $f_i$ are the same for all the hadrons.\\
The most interesting case is for $f \le 0$. As soon as $f$ moves from zero towards negative values the threshold energy at first slightly increases, but for $f<- 2.5 \cdot 10^{-14}~(n=1)$ [$f<- 3 \cdot 10^{-6}~(n=2)$],  eq.\ref{eq6} has no longer real solutions\cite{aletal}: the photo-pion production reaction is no longer kinematically allowed and protons propagate freely in the Universe.\\
For nuclei,  for which the relevant process of interaction on the universal backgrounds is photo-disintegration, an equation corresponding to eq.\ref{eq6}, with  $m_P \rightarrow A m_P$, can be written. The modification of the thresholds is similar to that for protons.\\
Limits on LIV parameters derived from the observed steepening of the spectrum of UHECRs have been reported in literature\cite{pliv}\cite{nliv}.\\
These limits,  however, depend crucially on the assumption that the observed flux suppression is originated by the propagation of UHECRs. 
Auger composition data combined with those on the all-particle spectrum might indicate a different scenario, as illustrated for example by Blasi\cite{blasi2014}. According to\cite{blasi2014} the two first moments of the distribution of $X_{max}$, the depth in atmosphere where the shower reaches its maximum development, may  indicate that the flux suppression is due to the end of cosmic ray acceleration at the source, implying also a very hard injection spectrum, incompatible with Fermi acceleration mechanism. In this framework propagation would have little, if any, effects on experimental observables.\\
 It is therefore worthwhile to verify if LIV  can be still bound in  this scenario.\\
\begin{figure}[!htb]
\centering
\includegraphics*[scale=0.3]{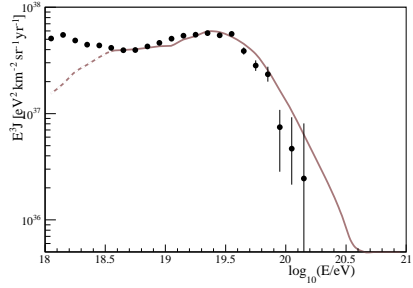} \hspace{0.5cm}
\caption{\it The all particle flux compared with the LIV case in the text.}
\label{fig1}
\end{figure}
\begin{figure}[!htb]
\centering
%\begin{tabular}{l}
\includegraphics*[scale=0.6]{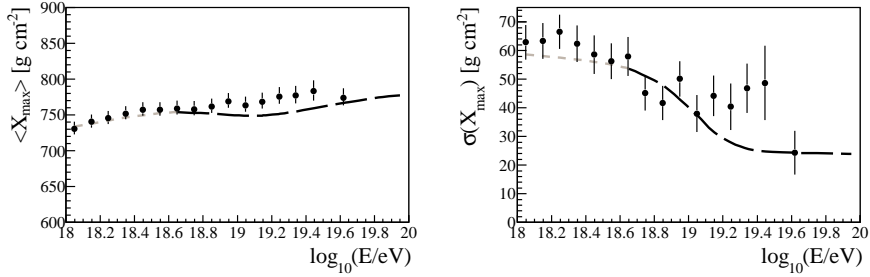}\\
%\includegraphics*[scale=0.6]{Xmaxgamma2Ecut20c.eps}\\
%\includegraphics*[scale=0.6]{lnAplots_ICRC2013.eps}\\
%\includegraphics*[scale=0.6]{lnAgamma2Ecut20.eps}\\
%\end{tabular}
\caption{\it $\langle X_{max} \rangle$  and its dispersion $\sigma( X_{max})$ as reconstructed from the LIV simulation described in this paper.}
\label{fig2}
\end{figure} 
To simulate LIVs we have  propagated  UHECRs switching off the interactions with background photons, only accounting for energy losses due to the expansion of the Universe. To account for these losses we used a simplified version of SimProp\cite{SimProp}, with maximum source rigidity $R_{max}=5 \cdot 10^{18} ~V$ and fixed $\gamma=2$, consistent with Fermi acceleration. The source model used is the minimal, ``standard'' one, $i.e.$ equal sources, uniformily distributed in comoving volume throughout the whole Universe, without evolution effects, emitting all nuclei in a rigidity dependent way.\\
The simulation is consistent with LIV as soon as $f$ is sufficiently negative so that Eq. (\ref{eq6}) has no real solution.   
The produced fluxes and composition qualitatively reproduce both Auger spectrum and composition behaviour as shown in fig.\ref{fig1} and fig.\ref{fig2}. \\
This has an important consequence: the present data from the Pierre Auger observatory, interpreted in the simple framework above\footnote{Note however that this framework, as also indicated for instance in\cite{abb} can only fit the data above $4 \cdot 10^{18}eV$ and a different component is needed at lower energies.}  do not allow to constrain LIV effects as parametrized by modified dispersion relations ( eq.\ref{eq3},\ref{eq6}).\\% and discussed in\cite{pliv}\cite{nliv}. \\
It is however obvious that the above statement $cannot$ be taken as evidence of LIV, since many other astrophysical/particle physics explanations can be considered. For instance the source model is too simple. On the other hand possible sources with hard spectrum have been proposed\cite{blasi2014}. Moreover, changes in the hadronic cross sections above LHC energies cannot be excluded, and would modify UHECR interactions.
Finally, for completeness, we note that the Telescope Array Collaboration has reported indications of a proton-dominated cosmic ray composition\cite{TA}. With the current statistics, Telescope Array data cannot discriminate between the proton and Auger-like composition\cite{PAOTA}. A proton composition would invalidate the conclusion that data are compatible with LIV, if the reported spectrum suppression is due to propagation.
\section{Lorentz Invariance Violations: other effects on UHECR Physics \label{sec.livacc}}
In principle, $all$ aspects of UHECR physics can be modified by LIV.\\
For instance, LIV can affect the cosmic ray acceleration processes, and  also the energy losses during acceleration. Since in the example of LIV propagation in the above section we considered $\gamma=2$, we can assume standard Fermi acceleration.\\ 
With respect to acceleration itself changes might be possible since (at UHE) $E \neq p$ due to LIV. However we already commented that this modification is very small and only relevant near the QG scale. Moreover, even in the case of relativistic shocks the Lorentz factor of the shock is much smaller than that of the accelerated particles, and therefore LIV effects on the shock itself are not expected. \\
% For a different approach to LIV effects in acceleration see for instance\cite{RK}. \\
For (synchrotron) energy losses at the source there might be a more important effect because, since ($f_i<0$) the group velocity of nuclei reaches a maximum value $<c$, the Lorentz factor of, say, a proton is bounded  and the energy lost in photons is limited\cite{libetal2002}.
This point will be discussed in a forthcoming paper.\\
More important effects are expected in the interactions of UHE particles in the atmosphere and in the decay of secondary particles. These effects can make some parts of the kinematical space unallowed for the processes and therefore make some reaction impossible. 
With respect to the modification of the thresholds discussed in the previous Section, an important difference is that these processes might be affected by (unknown) LIV dynamics. However, since we are interested in $conservative$ bounds, we do not consider this problem here. In the next two subsections we will discuss, in a unified kinematical approach, the effects on particle decays, hence atmospheric showering, and interactions of nuclei in the atmosphere.
\subsection{LIV effects on particle decays and showering}
\begin{figure}[!htb]
\centering
\begin{tabular}{l}
\includegraphics*[width=0.75\textwidth]{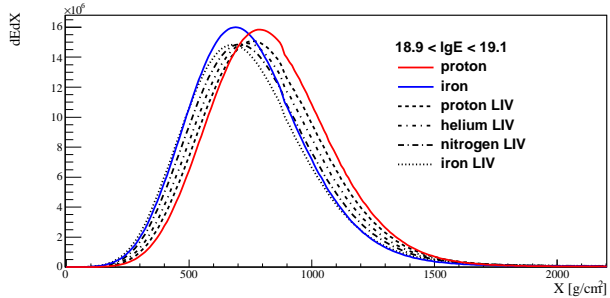}\\
\end{tabular}
\caption {\it Extensive Air Shower longitudinal development simulated with CONEX. Red and blue solid line represents the case of standard LI shower development respectively for protons and iron primaries. The dashed lines represent the LIV cases for  different masses. }
\label{fig4}
\end{figure}

\begin{figure}[!htb]
\centering
\begin{minipage}{.5\textwidth}
  \centering
  \includegraphics[width=1\linewidth]{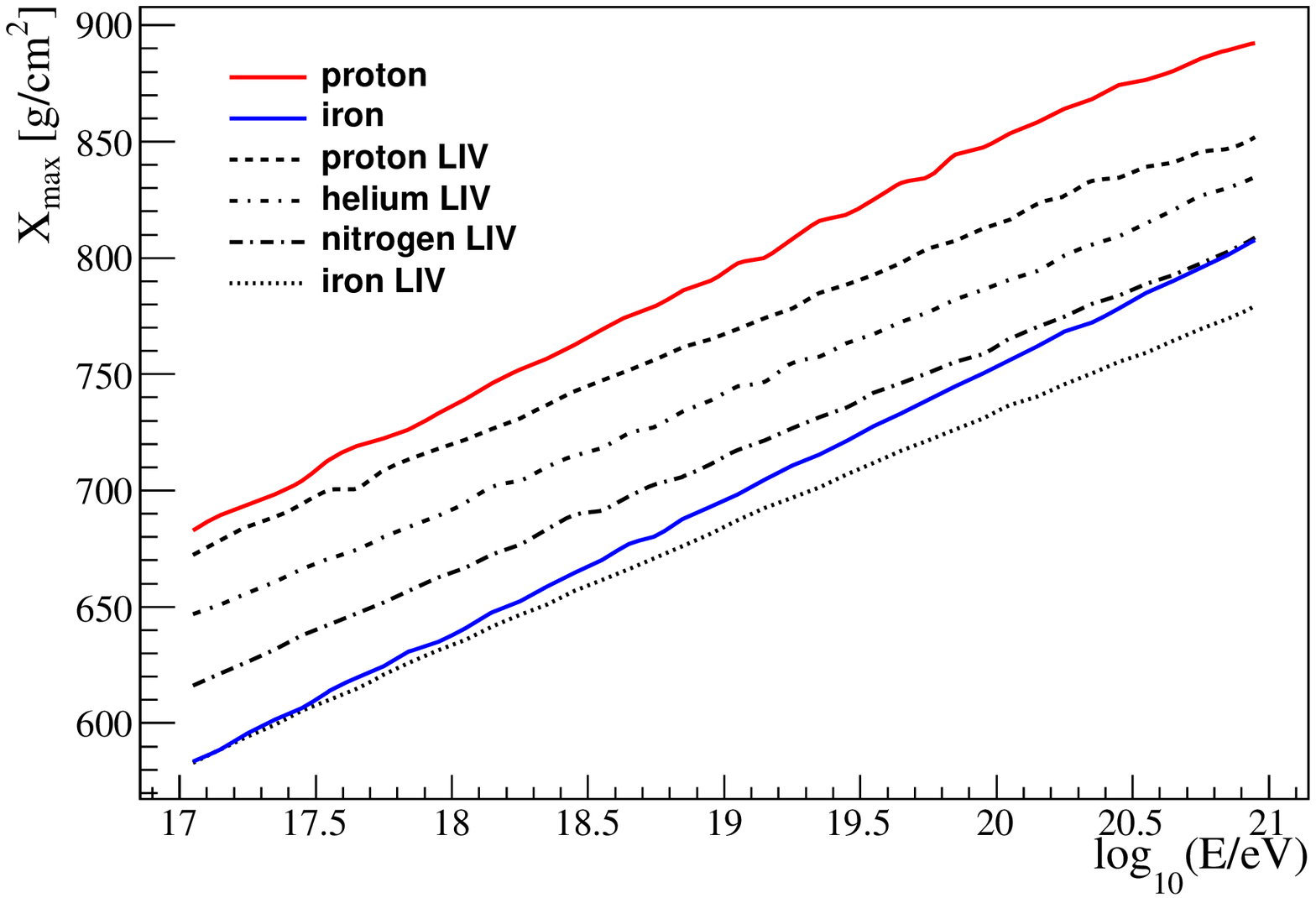}
%  \label{fig:test1}
\end{minipage}%
\begin{minipage}{.5\textwidth}
  \centering
  \includegraphics[width=1\linewidth]{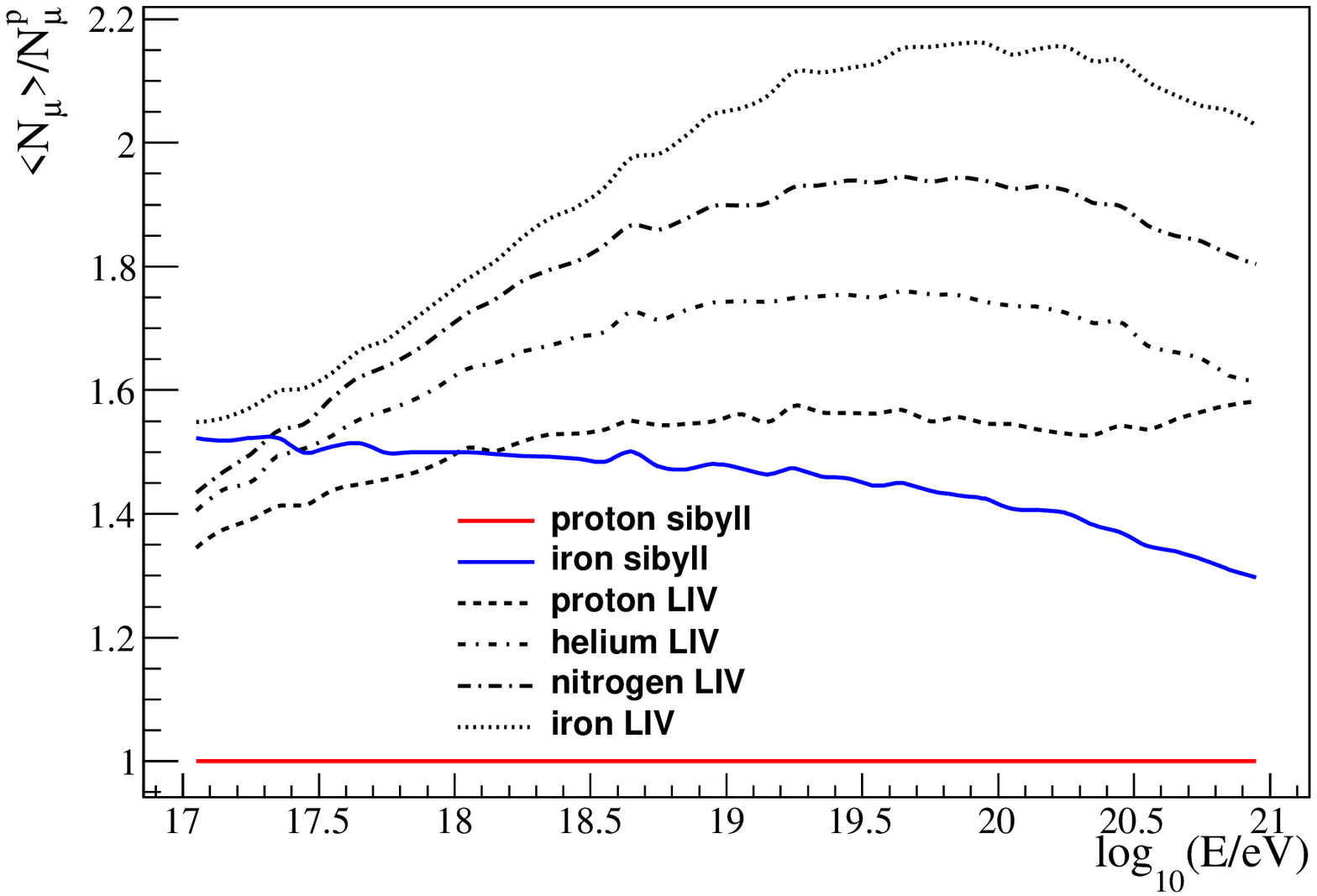}
%  \label{fig:test2}
\end{minipage}
\caption{\it Left panel: Expectation for $\langle X_{max} \rangle $ vs energy for LI shower development (solid lines) and LIV case (dashed lines). The LIV effect at the  highest energies is to make shower appear heavier than they are in reality. Right panel: Average number of muons vs primary energy in LI and LIV cases. This number has been normalized to the average amount of muons in standard LI proton showers to better show the effect of LIV.}
  \label{fig44}
\end{figure}

Consider the most important decay for atmospheric showering, $\pi^0 \rightarrow \gamma \gamma $. We construct, both for the initial particle and the final state,  the quantity $s=( \sum p_i^{\mu})^2$. When $f=0$ (LI) $s$ is an invariant and can be computed in any reference frame; if $f\neq 0$ (LIV) this is not the case but energy-momentum conservation implies that this quantity should be equal between  initial and final states, if computed in the same reference frame. Now the crucial point is that, with $f<0$ there is no guarantee that this quantity is still positive. For the above decay, from the equality $s_{ini}=s_{fin}$ we obtain:
\begin{equation}
m^2_{\pi}+{\frac{1}{m_P^n}}(f_{\pi}E^{2+n}_{\pi}-f_{\gamma}(E_{\gamma_1}^{2+n}+E_{\gamma_2}^{2+n})) - 2 (E_{\gamma_1}E_{\gamma_2}-p_{\gamma_1} p_{\gamma_2}) =2 p_{\gamma_1} p_{\gamma_2} (1-\cos \theta_{1,2})
\label{eq12}
\end{equation}
Since there are very strong limits\cite{stanetal} on $f_{\gamma}$ we will assume it to be zero.
The right hand side of eq.\ref{eq12} is non negative, while the left hand one can become negative for large enough $E_{\pi}$. Therefore neutral pions do not decay if
$ E_{\pi}>(m_P^n m_{\pi}^2/|f_{\pi}|)^{\frac{1}{2+n}}$.
To test this effect we have generated $100000$ atmospheric showers with CONEX\cite{CONEX} imposing the same condition for all particle decays. The results of this simulation are presented in figs.\ref{fig4},\ref{fig44}. In particular, in fig.\ref{fig4} the air shower longitudinal development,
in the case of standard LI development (for protons and iron
primaries), is compared to the LIV case for different masses. Since the  energy of the pions is related to the energy per nucleon of the incident nucleus, the LIV threshold moves to higher energies for heavier nuclei.\\
In fig.\ref{fig44}, left panel, the expectation for $\langle X_{max} \rangle $ vs energy for LI
shower development (solid lines) and LIV case (dashed lines) is reported, while  the
right panel  presents the average number of muons vs primary energy in LI and LIV cases.
This number has been normalized to the average number of muons in standard
LI proton showers to better show the effect of LIV. \\
The suppression of the (neutral) pion decays makes these particles interact, thus increasing the amount of muons in the extensive air shower.
Moreover the position of the shower maximum  moves to higher altitudes as the electromagnetic part of the shower consumes faster.
 From the observational point of view this makes nuclei (and protons) primaries looking heavier than they are in reality.
These changes in the shower development will also affect the results
reported in the previous section, since the knowledge of the shower developement is a necessary ingredient
to perform the comparison with experimental data.
Detailed study is underway and will be presented in a further publication.\\
\subsection{LIV effects on interactions}
The interactions of UHECR nuclei can also be affected by LIV.
To discuss these effects, we follow here the same approach of the previous subsection, and consider for instance the reaction $ p_{CR} p_{air} \rightarrow p_1 p_2+n_{\pi} \pi$.
\begin{eqnarray}
s_{ini} & = & (p_{CR}^{\mu}+p_{Air})^2 = 2m^2+2(p_{CR}^{\mu}p_{Air}^{\mu})+{f \over m_p} E^3 \nonumber \\
        &  \approx & 2 m^2+2 E_{CR}m+{f \over m_p} E_{CR}^3 
\label{eq8}
\end{eqnarray}
having neglected LIV for the nucleons of the atmospheric nuclei. If $ f < 0$ and $ E_{CR}>\sqrt{2 m (m+m_P)/(-f)} \approx 5\cdot 10^{18}~eV$ ($f=-1$) then $s_{ini}<0$. \\
Of course there can be cancellations, since also in the right of the reaction there will be (negative) LIV terms. However, given the energy dependence of the LIV term, an exact cancellation is only possible in the elastic case ($n_{\pi}=0$) and if the CR proton does not lose energy.\\
The equality  $s_{ini}=s_{fin}$ implies, in the case $n_{\pi}=0$ taken as an example:
\begin{equation}
 2 m E_{CR}+{f \over m_p}( E_{CR}^3 - E_1^3 -E_2^3 ) 
 -  2 (E_1  E_2-p_1  p_2)  =  
 2(p_1  p_2(1- \cos \theta_{12}))  
\label{eq10}
\end{equation} 
%\begin{eqnarray}
%&    & 2 m E_{CR}-n m_{\pi}^2+{f \over m_p}( E_{CR}^3 - E_1^3 -E_2^3-\sum E_{\pi,i}^3 ) \nonumber \\  
%& - & 2 (E_1  E_2-p_1  p_2)-2 \sum (E_1  E_{\pi,i}-p_1  p_{\pi,i}) \nonumber \\ 
%& - & 2 \sum (E_2  E_{\pi,i}-p_2  p_{\pi,i})-2\sum(E_{\pi,i}E_{\pi,j}-p_{\pi,i}  p_{\pi,j}) =  \\ 
%&    & 2(p_1  p_2(1- \cos \theta_{12}))+2\sum (p_1  p_{\pi,i}(1- \cos \theta_{1i}))  \nonumber \\
%& + & 2\sum (p_2  p_{\pi,i}(1- \cos \theta_{2i})) +2\sum (p_{\pi,i}p_{\pi,j}(1- \cos \theta_{ij})) \nonumber \label{eq10}
%\end{eqnarray} 

Again the right hand side   of eq.\ref{eq10} is non negative by construction. On the other hand, if $ f<0$ the left hand side can be negative for large enough $E_{CR}$.
Numerically one finds that as soon as $E_{CR} \geq 10^{19}~eV,~(f=-1)$, the left hand side becomes negative apart in a very small kinematical region so that the reaction is not allowed.
This means for instance that if we clearly detect at ground (the interaction of) a proton with primary energy of $E=10^{20}~eV $ we can set a limit for   $f \geq -5 \cdot 10^{-3}$. \\
These effects will also affect the shower developement. As above, only detailed simulation can describe the overall effect.
\section{Conclusions \label{sec.conc}}
In this note we have presented a discussion on the status of bounds on Lorentz Invariance Violations parameters at the light of most recent spectrum and composition data from the Pierre Auger Observatory\cite{fluxpao,ICRC2013}. If the data are interpreted as indicating that the spectrum of UHECRs is limited at the sources\cite{abb}, it turns out that the very strong limits that were previously derived\cite{pliv,nliv}, from the presence of the GZK flux suppression, do not apply any longer.
This does not affect other limits, derived from the mere existence of UHECRs \cite{RK}.\\
Clearly this fact cannot be interpreted as evidence for LIV since there are possible astrophysical/particle physics  explanations of the data.\\
We have then analyzed other aspects of UHECR physics that can be affected by LIV, in particular effects on interaction on the atmosphere and shower developement: LI violating interactions and decay can induce modifications of the normal physics which dictates the production of secondary particles that are detected in UHECRs experiments. The effects of these  modifications are in principle detectable (and falsifiable) in an experiment like the Pierre Auger Observatory: in order to understand if this can be done effectively, however, detailed simulations are needed and are under way.
\section{Acknowledgements}We would like to acknowledge many useful discussions with our colleagues of the Pierre Auger Collaboration. We also thank G. Amelino-Camelia, S.Liberati and F.Mendez for discussions in an early part of this work, and T. Pierog for help in modifying CONEX.

\end{document}